\documentclass{article}
\usepackage{ijcai19}

\usepackage{times}
\usepackage{soul}
\usepackage{url}
\usepackage[hidelinks]{hyperref}
\usepackage[utf8]{inputenc}
\usepackage[small]{caption}
\usepackage{graphicx}
\usepackage{amsmath}
\usepackage{booktabs}
\usepackage{algorithm}
\usepackage{algorithmic}
\usepackage{multirow}
\usepackage{booktabs} 
\usepackage{enumerate}
\usepackage{subfigure}
\usepackage{epstopdf}
\usepackage{balance}
\usepackage{flushend}
\usepackage{multirow}
\usepackage{enumerate}
\usepackage{caption}
\usepackage{verbatim}

\newtheorem{defn}{Definition}[section]

\title{Fine-grained Event Categorization with Heterogeneous Graph Convolutional Networks}
\author{
Hao Peng$^{1,2}$,
Jianxin Li$^{1,2}$,
Qiran Gong$^{1,2}$,
Yangqiu Song$^{3}$,
Yuanxing Ning$^{1,2}$,\\
Kunfeng Lai$^{4}$ \normalfont{and} \textbf{Philip S. Yu}$^{5,6}$
\affiliations
$^1$ Beijing Advanced Innovation Center for Big Data and Brain Computing, Beihang University\\
$^2$ State Key Laboratory of Software Development Environment, Beihang University\\
$^3$ Department of Computer Science and Engineering, Hong Kong University of Science and Technology\\
$^4$ Platform and Content Group, Tencent\\
$^5$ Institute for Data Science, Tsinghua University\\
$^6$ Department of Computer Science, University of Illinois at Chicago\\
\emails
\{penghao, lijx, ningyx\}@act.buaa.edu.cn, allen\_gong@buaa.edu.cn, yqsong@cse.ust.hk,\\ calvinlai@tencent.com, psyu@uic.edu
}

\begin{document}

\maketitle

\begin{abstract}
Events are happening in real-world and real-time, which can be planned and organized occasions involving multiple people and objects.
Social media platforms publish a lot of text messages containing public events with comprehensive topics.
However, mining social events is challenging due to the heterogeneous event elements in texts and explicit and implicit social network structures.
In this paper, we design an event meta-schema to characterize the semantic relatedness of social events and build an event-based heterogeneous information network (HIN) integrating information from external knowledge base, and propose a novel Pairwise Popularity Graph Convolutional Network (PP-GCN) based fine-grained social event categorization model.
We propose a Knowledgeable meta-paths Instances based social Event Similarity (KIES) between events and build a weighted adjacent matrix as input to the PP-GCN model.
Comprehensive experiments on real data collections are conducted to compare various social event detection and clustering tasks.
Experimental results demonstrate that our proposed framework outperforms other alternative social event categorization techniques.
\end{abstract}

\section{Introduction} 
Events are happening in real-world and real-time, which can be planned and organized occasions involving multiple people and objects, such as a social gathering, celebrity activities or a sports competition in some specific location at a particular time.
Nowadays, social media platforms have become major sources for publicizing events.
Events announced on social media usually attract comments and reposts with opinions and emotions, and such content can reflect public opinion about many social, political, economic issues, etc.
Mining of social media posts, such as fine-grained social event categorization, will benefit a lot of real applications, such as information organization, predictive analysis, disaster risk analysis, and others~\cite{atefeh2015survey,aggarwal2012event,Allan:2012:TDT:2481012}.
In general, fine-grained social event categorization focus on event detection and event clustering.

The tasks of fine-grained social event categorization are more challenging than traditional text mining or social network mining, since social event is a combination of social network and the information flows (in terms of short messages) over it.
On the one hand, modeling social events is very complicated and ambiguous.
Social events are described in short texts and usually contain different types of entities, such as person, location, organization, number, time, etc~\cite{Allan:2012:TDT:2481012,ji2008refining,yu2017ring}.
Moreover, events are commented or retweeted by social network users.
Thus, modeling social event needs to consider heterogeneous elements as well as explicit and implicit social network structures within social posts.
On the other hand, models of fine-grained event categorization often have bottlenecks in which the number of the categories is large and the number of samples per class is small.
Thus, fine-grained event categorization needs to address the accuracy of the developed algorithms.
Currently, fine-grained text classification is more difficult and lacks related research work than the fine-grained object recognition in other fields such as computer vision~\cite{zhang2014part}.

A handful of studies~\cite{ritter2012open,chandola2009anomaly,becker2011identification,shao2017efficient} have investigated leveraging homogeneous graphs or manually defined frames for social event modeling and extracting.
The first line of thought is to treat social event as homogeneous words/elements co-occurrence graph~\cite{chandola2009anomaly,aggarwal2012event,angel2012dense,liu2019event}.
Typically, they construct a homogeneous words/elements co-occurrence graph, and then consider different scales of abnormally connected subgraph structures (under different names such as k-clique, motifs or graphlets) as the social events.
Despite the compelling results achieved by these studies, their categorization accuracies remain unsatisfactory for building reliable and open domain event detection and clustering systems in practice.
The second line of thought is to use manually defined frames-based event definitions applying the well-defined techniques for extracting social event frames from news~\cite{kim2009overview,ji2008refining}.
The frame-based event extraction can extract entities and their relationships, but uses only a limited number of event types, such as earthquake disaster, stock market, venues, politics, etc.
Moreover, it uses complicated machine learning models, usually a pipeline of them, to incorporate different levels of annotation and features.

Social media events can be regarded as a co-occurrence of event elements including themes, dates, locations, people, organizations, keywords and social behavior participants.
The simplest way to monitor social media events is to represent events as bags-of-words, but it will be more semantically meaningful if we can annotate words and multi-word-expressions as entities with types.
For example, in the tweet "\emph{China Seismological Network: The earthquake struck at 21:19:46 China Standard Time on 8 August 2017 in Zhangzha Town in Jiuzhaigou County with magnitude 7.0}", there are multiple event elements: \textbf{Time}: \emph{21:19:46}; \textbf{Date}: \emph{August 8, 2017}; \textbf{Timezone}: \emph{China Standard Time}; \textbf{Town}: \emph{Zhangzha}; \textbf{County}: \emph{Jiuzhaigou}; \textbf{Nation}: \emph{China}; \textbf{Magnitude}: \emph{7.0}; \textbf{Poster}: \emph{China Seismological Network}.
Obviously, the above event's elements are of different types. 
Moreover, in addition to intuitive co-occurrence relationship, after extracting entities, we can make use of external knowledge base~\cite{Auer:2007:DNW:1785162.1785216,xu2017cn} to complement more relationships between entities, such as “\textbf{located-in}” relationships with other locations, “\textbf{attribute-of}” relationships with magnitude and earthquake, etc.
Thus, a message mentioning an event can be related to its keywords, entities (and their relations), topics, etc.
Furthermore, the social network users posting messages are also connected with different relationships, such as following/followed and retweeting. 
Thus we can model social media events as HIN~\cite{shi2017survey}.

In this paper, we first present event instance (shown in short text message) as hyper-edge in an HIN, where all the keywords, entities, topics and social users can be connected by this hyper-edge, and define an event meta-schema to characterize the semantic relatedness of social event instances and build event-based HIN.
In order to enrich the HIN, we extract some information as a complement of the relationships based on the external knowledge base and algorithms.
Based on the event HIN, we define a weighted \underline{K}nowledgeable meta-paths \underline{I}nstances based \underline{E}vent \underline{S}imilarity measure, namely KIES, from semantically meaningful meta-paths.
In order to accurately measure the weights between meta-paths and perform fine-grained event detection, we then design a novel \underline{P}airwise \underline{P}opularity \underline{G}raph \underline{C}onvolutional \underline{N}etwork model, namely PP-GCN, to learn the representation of each event instance.
Finally, under the HINs-based event modeling, we present a KIES-measure based fine-grained event clustering.

Compared to traditional methods, the proposed models have several advantages:
(1) By modeling social events based on a HIN, the proposed framework can integrate event elements, such as keywords, topic, entities, social users and their relations, in a semantically meaningful way, and can also calculate the similarity between any two event instances.
(2) By modeling pairwise popularity graph convolutional network, the model achieves state-of-the-art results and avoids overfitting in fine-grained event detection tasks.
(3) The proposed KIES with learned weights between meta-paths by the PP-GCN can boost the performance of fine-grained social events clustering compared to existing state-of-the-art baselines methods.
The code of this work is publicly available at \emph{https://github.com/RingBDStack/PPGCN}.

\section{Heterogeneous Event Modeling}
In this section, we define the problem of modeling social events in heterogeneous information network (HIN) and introduce several related concepts and necessary notations.

\subsection{Event Modeling in HIN}
The definition and characterization of ``social event'' have received substantial attention across academic fields, from language~\cite{miller1998wordnet} to cognitive psychology~\cite{zacks2001event}. 
A social event generally refers to influential facts that appear on social networks and occur in the real world, including creators (posters), named entities such as participants, organizations, festival, specific times, places, currency, address, etc., and other elements such as keywords and topics.
We name the above elements as \emph{event-oriented elements}.
However, extracting the {event-oriented elements} from the original social text message with NLP tools\footnote{https://github.com/stanfordnlp/CoreNLP},\footnote{https://github.com/NLPIR-team/NLPIR} is still a prior processing work.
Even within most of the events, there are some relationships between {event-oriented elements}, such as relationships between entities, relationships between keywords, relationships between topics, explicit and implicit relationships between social users, and so on.
We name the above relationships as \emph{event-elements relationships}.

\begin{figure*}[t]\label{fig:hin}
\begin{center}
    \subfigure[Example of two event instances connected by different types of nodes and edges ]{\label{fig:event-hin}
\includegraphics[width=0.50\textwidth,height=0.20\textheight]{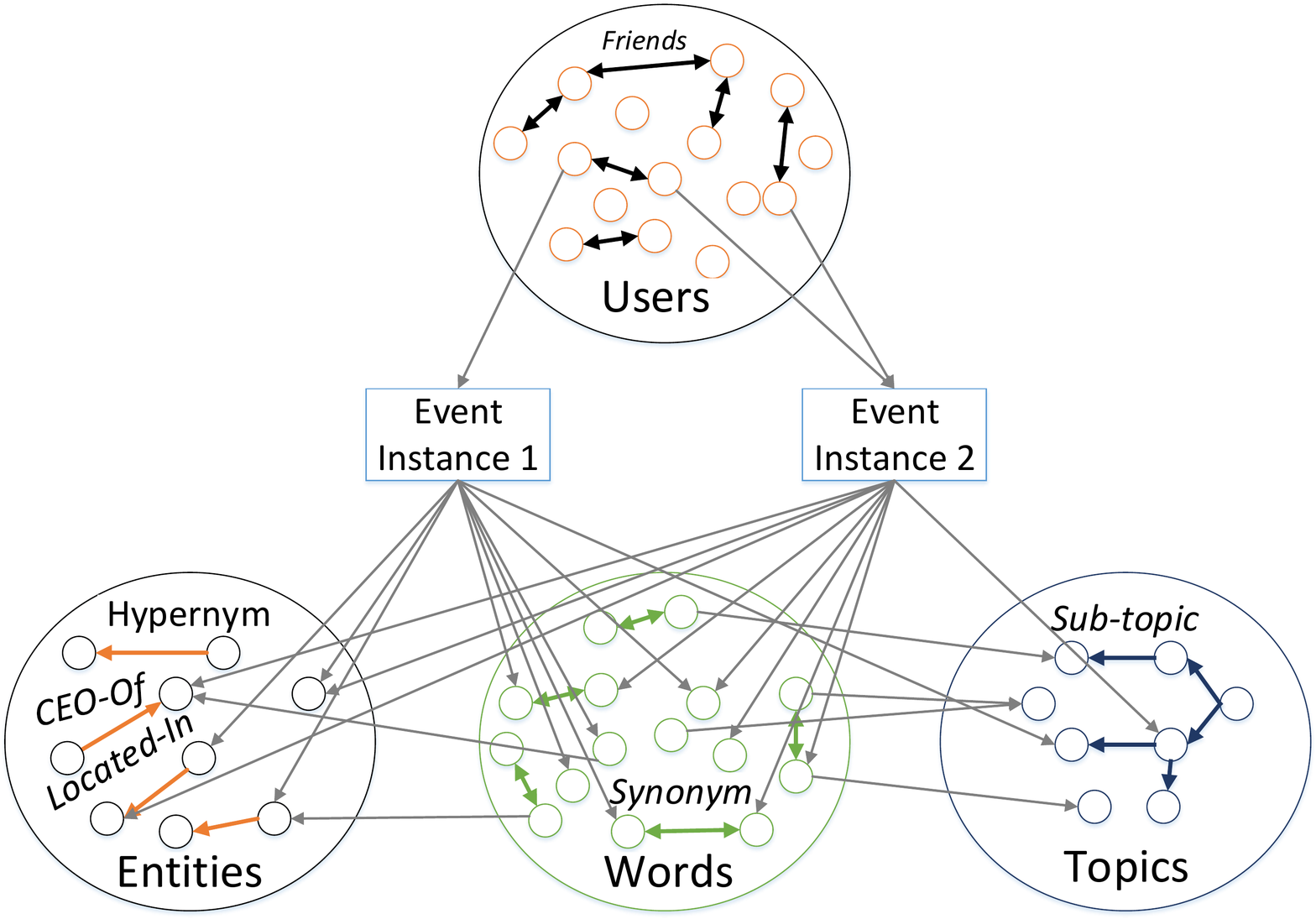}
	}
	\subfigure[Meta-schema of event-based HIN]{\label{fig:event-schema}
\includegraphics[width=0.45\textwidth,height=0.20\textheight]{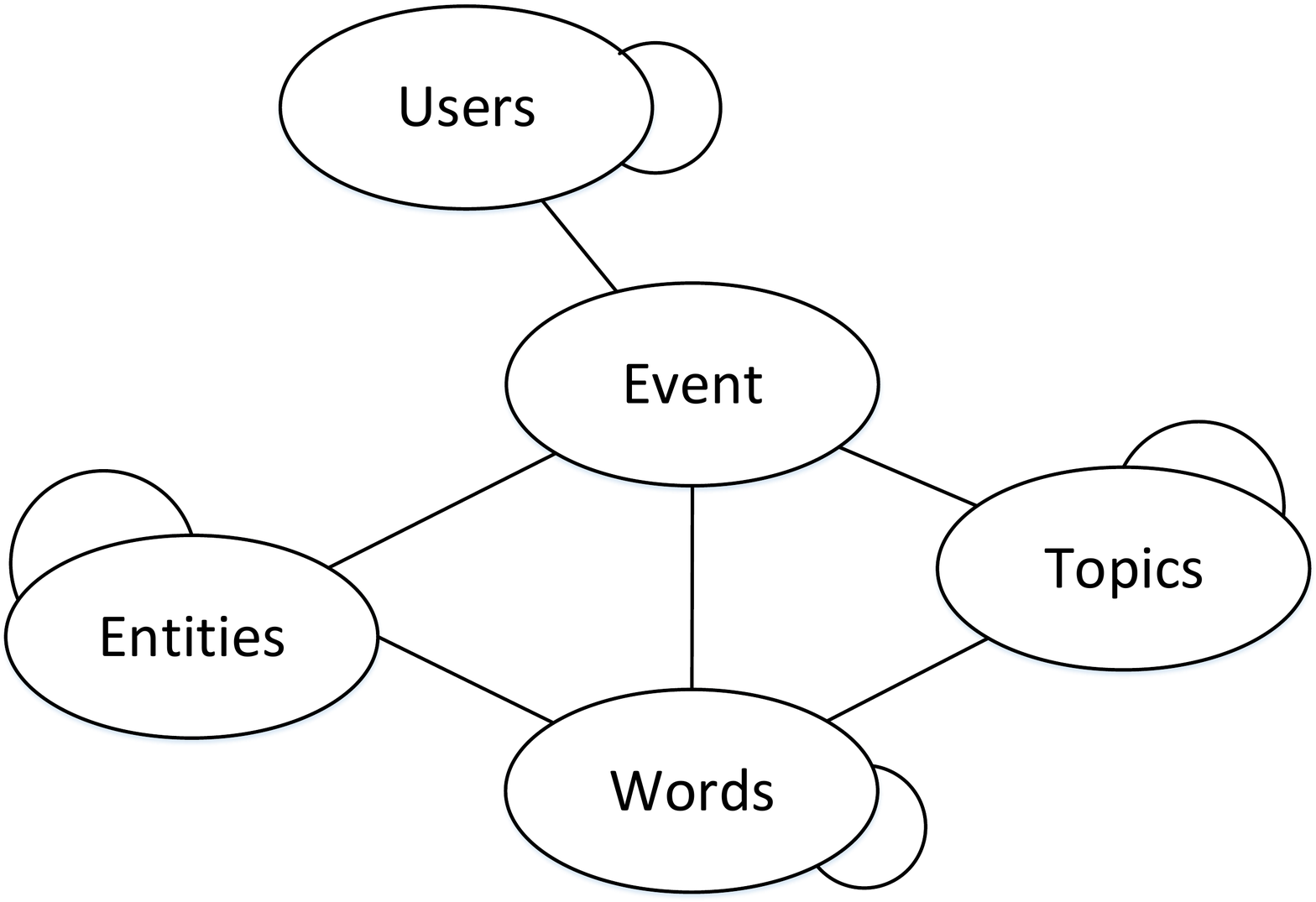}
	}
\end{center}
	\caption{Illustration of events as a HIN, where event instances are represented as hyper-edges.}
\end{figure*}

We use the manually organized synonyms\footnote{https://github.com/huyingxi/Synonyms} to add synonym relationship among keywords in the event-based HIN.
For hierarchical topic structures and the affiliation relationship between keywords and topics in the event-based HIN, we employ the hierarchical latent Dirichlet allocation technologies~\cite{griffiths2004hierarchical,blei2010nested} based on the existing toolbox\footnote{https://github.com/joewandy/hlda} (with about 30 most probable words for each topic).
In order to build the relationship between entities in the event-based HIN, we consider both accuracy and efficiency, and tackle the problem by following three-steps.
First, we retrieve the same entity candidate from knowledge base, such as the Chinses CN-DBpedia~\cite{xu2017cn}.
Second, we use word embeddings~\cite{mikolov2013distributed} based Word Mover's Distance technology~\cite{kusner2015word} to measure the similarity between context of entity in the social text and description of entity in the candidate, and choose the entity from the candidate with highest similarity.
Third, we query the relationship between aligned entities in the knowledge base as the final relationship of entities in \emph{event-elements relationships}.
In order to establish the relationship between entities and keywords in the event-based HIN, we extract the keywords in the relevant description of each entity in the knowledge base and use this affiliation as the relationship between the entity and the keywords.
For the relationship between social users, we consider users with a large number of friends, and store the relationship between users in advance.

After extracting the above {event-oriented elements} and {event-elements relationships} from event instances, we build an event-based HIN, as shown in Figure~\ref{fig:event-hin}.
The social event can be regarded as a co-occurrence of {event-oriented elements}, and the {event-elements relationships} are conducive to explaining the relationship between various elements.
Thus, an event instance can be treated as a subgraph of the whole HIN. 
One particular advantage of the HIN is that meta-paths defined over types (e.g., a typical meta-path ``event-entity-event'' represents the event similarity based on overlapped entities between two event instances) can reflect semantically meaningful information about similarities, and thus can naturally provide explainable results for event modeling.

\subsection{Preliminaries}
We introduce some basic definitions from previous works~\cite{sun2013mining,shi2017survey}, and give some event-HIN examples.
\begin{defn}\label{def:hin}
A \textbf{heterogeneous information network} (HIN) is a graph $G = (V,E)$ with an entity type mapping $\phi: V\to A$ and a relation type mapping $\psi: E\to R$, where $V$ denotes the entity set, $E$ denotes the link set, $R$ denotes the relation type set and $A$ denotes the entity type set. The number of entity types $|A|>1$ or the number of relation types $|R|>1$.
\end{defn}
For example, Figure~\ref{fig:event-hin} shows an example of two event instances connected with different types of entities, keywords, topics, social users and relationships.
After giving a complex HIN for event modeling, it is necessary to provide its meta level (i.e., schema-level) description for better understanding. 
\begin{defn}\label{def:meta-schema}
Given an HIN $G = (V,E)$ with the entity mapping $\phi: V\to A$ and the relation type mapping $\psi: E\to R$, the \textbf{meta-schema} (or network schema) for network $G$, denoted as $T_{G} = (A,R)$, is a graph with nodes as entity types from $A$ and edges as relation types from $R$.
\end{defn}
For example, Figure~\ref{fig:event-schema} shows an example of the HIN meta-schema characterizing events on social messages.
Another important concept is the meta-path which systematically defines relationships between entities at the schema level.
\begin{defn}\label{def:meta-path}
A \textbf{meta-path} P is a path defined on the graph of network schema $T_{G} = (A,R)$ of the form $A_{I}\stackrel{R_{1}}{\longrightarrow}A_{2}\stackrel{R_{2}}{\longrightarrow}A_{3}\cdots A_{L}\stackrel{R_{L}}{\longrightarrow}A_{L+1}$ which defines a composite relation $R=R_{1}\cdot R_{2}\cdot \cdots \cdot R_{L}$ between objects $A_{1},A_{2},A_{3}\cdots A_{L+1}$, where $\cdot$ denotes relation composition operator, and $L+1$ is the length of $P$.
\end{defn}
For simplicity, we use object types connected by $\to$ to denote the meta-path when there are no multiple relations between a pair of types: $P = (A_{1} - A_{2} - \cdots - A_{L+1})$.
We say that a meta-path instance $p = (v_{1} - v_{2} - \cdots -  v_{L+1})$ between $v_{1}$ and $v_{L+1}$ in network $G$ follows the meta-path $P$, if $\forall l$, $\phi(v_{l}) = A_{l}$ and each edge $e_{l} = <v_{l}, v_{l+1}>$ belongs to each relation type $R_{l}\in P$.
We call these paths as path instances of $P$, denoted as $p\in P$.
$R_{l}^{-1}$ represents the reverse order of relation $R_{l}$.
We will introduce more semantically meaningful meta-paths that describe event relations in next section.

\section{The Proposed Model} 
In this section, we introduce definitions about knowledgeable meta-paths instances based event similarity measure, and present the technical details about Pairwise Popularity GCN.

\subsection{Event Similarity Measure}
Before definite the social event similarity, we first present the definition of \emph{CouP} as following,
\begin{defn}\label{def:countsim}
\textbf{CouP}: Given a meta-path $P = (A_{1} - A_{2} \cdots A_{L+1})$, CouP is a function of the count of meta-path instances such that $CouP_{P}(v_{i},v_{j}) = M_{P}(v_{i},v_{j})$ where $M_{P}=W_{A_{1}A_{2}}\cdot W_{A_{2}A_{3}}\cdots W_{A_{L}A_{L+1}}$ and $W_{A_{k}A_{k+1}}$ is the adjacency matrix between types $A_{k}$ and $A_{k+1}$ in the meta-path $P$.
\end{defn}
For example, for event instance similarity based on {event-oriented elements} and {event-element relationships}, the composite relation of two event instances containing the same {event element} and co-occurrence relationship can be described as "event Instance - Element - event Instance (IEI)" for simplicity. 
This meta-path simply gives us $M_{IEI} = W_{IE}W_{EI}^{T}$, which is the dot product between event instances, where $W_{EI}$ is the event Instance-Element co-occurrence matrix. 
The similarity based on this meta-path instances is accurate because different elements and lengths are considered. 
The more meta-paths enumerated by the meta-schema, the higher accuracy of the similarity metric is.
We can give more event related meta-paths over different lengths, e.g.,
$P_1$: \emph{Event-(posted by)-Social user-(post)-Event},
$P_2$: \emph{Event-(having)-Washington DC-(capital of)-United States-(contained by)-Event},
$P_3$: \emph{Event-(belong to)-Politician-(relevant)-president-(member of)-Ruling Party-(contained by)-Event}, etc.
$P_1$ means two event instances are similar if they are posted by the same social user.
$P_2$ means two event instances are similar if they mention Washington DC and the United States, respectively, where Washington DC is the capital of the United States.
$P_3$ means two event instances are similar if they can be associated by a chain of three event elements with meaningful relationships.
Note that the meta-path does not need to satisfy symmetry.
Here, we enumerate $22$ symmetric meta-paths in the meta-schema of event-based HIN.

However, if the counts are not normalized for different meta-paths, it is difficult to compare over different meta-path-based similarities. 
Then, similar to the HIN-based document similarity~\cite{wang2018unsupervised}, we also define our knowledgeable meta-paths instances based social event similarity measure, namely KIES.
Intuitively, if two event instances are more strongly connected by the important (i.e., highly weighted) meta-paths, they tend to be more similar.
\begin{defn}\label{def:kies}
\textbf{KIES: a knowledgeable meta-paths instances based social event similarity}. Given a collection of meaningful meta-paths, denoted as $\textbf{P}=\{P_{m}\}_{m=1}^{M'}$, the KIES between two event instances $e_i$ and $e_j$ is defined as: 
\begin{equation}\label{eq:event_sim_inst}\small 
KIES(e_i,e_j) = \sum_{m=1}^{M'}\omega_{m}\frac{2\times CouP_{P_m}(e_{i},e_{j})}{CouP_{P_m}(e_{i},e_{i})+CouP_{P_m}(e_{j},e_{j})},
\end{equation}
\end{defn}
where $CouP_{P_m}(e_{i},e_{j})$ is a count of meta-path $P_m$ between event instances $e_i$ and $e_j$, $CouP_{P_m}(e_{i},e_{i})$ is that between event instances $e_i$ and $e_i$, and $CouP_{P_m}(e_{j},e_{j})$ is that between event instances $e_j$ and $e_j$.
We use a parameter vector $\vec{\omega} = [\omega_{1},\omega_{2},\dots,\omega_{M'}]$ to denote the meta-path weights, where $\omega_{m}$ is the weight of meta-path $P_{m}$.
$KIES(e_i,e_j)$ is defined in two parts: 
(1) the semantic overlap in the numerator, which is defined by the number of meta-paths between event instances $e_i$ and $e_j$;
and (2) the semantic broadness in the denominator, which is defined by the number of total meta-paths between themselves.
Therefore, we can give a KIES distance with weights for any two event instances.

\subsection{Pairwise Popularity GCN Model}
Next, we show how to implement fine-grained event detection on social message texts through the pairwise popularity GCN model (PP-GCN), and learn the weights $\vec{\omega}$ for meta-paths, to overcome the problems of a large number of categories and a small number of samples per class.

After computing the distance of any two event instances by the KIES, we can construct a $N\times N$ weighed adjacent matrix $A$ for manually annotated social event instances, where $N$ is the number of event instances and $A_{ij} = A_{ji} = KIES(e_i,e_j)$.
Then, we train the Doc2vec~\cite{le2014distributed} representation as generalized event instance feature.
So, we can construct a $N\times d$ feature matrix $X$, where $d$ is the dimension of event instance feature.
Obviously, so far, we can use the popular GCN~\cite{kipf2017semi} architecture to learn discriminating event representation based on the interactions among event instances and generalized event instance features in node classification task.
The input to the GCN model includes the $A$ and $X$ matrices.
Here, one class represents one social event class.
In order to construct preliminary GCN model, we utilize the popular multi-layer GCN with the following layer-wise propagation rule~\cite{kipf2017semi}: 
\begin{equation}\label{eq:obj_gcn}
H^{(l+1)} = \sigma(\widetilde{D}^{-\frac{1}{2}}\widetilde{A}\widetilde{D}^{-\frac{1}{2}}H^{(l)}W^{(l)}),
\end{equation}
where $\widetilde{A} = A + I_{N}$, $\widetilde{D}$ is diagonal matrix such that $\widetilde{D}_{ii}=\sum_{j}\widetilde{A}_{ij}$ are the adjacency matrix, $I_{N}$ is the identity matrix, $W$ is the parameter matrix, and $l$ is the number of layers.
Let $Z$ be an output $N\times F$ feature matrix, where $F$ is the dimension of output representation per event instance.
The input layer to the GCN is $H^{(0)} = X, X\in R^{N\times d}$, which contains original event instance feature, $H^{(l)} =Z$, and $Z$ is graph-level output.
And $\sigma$ denotes an activation function such as Sigmoid or ReLU.

\begin{figure}[t]
\centering
\includegraphics[width=0.5\textwidth]{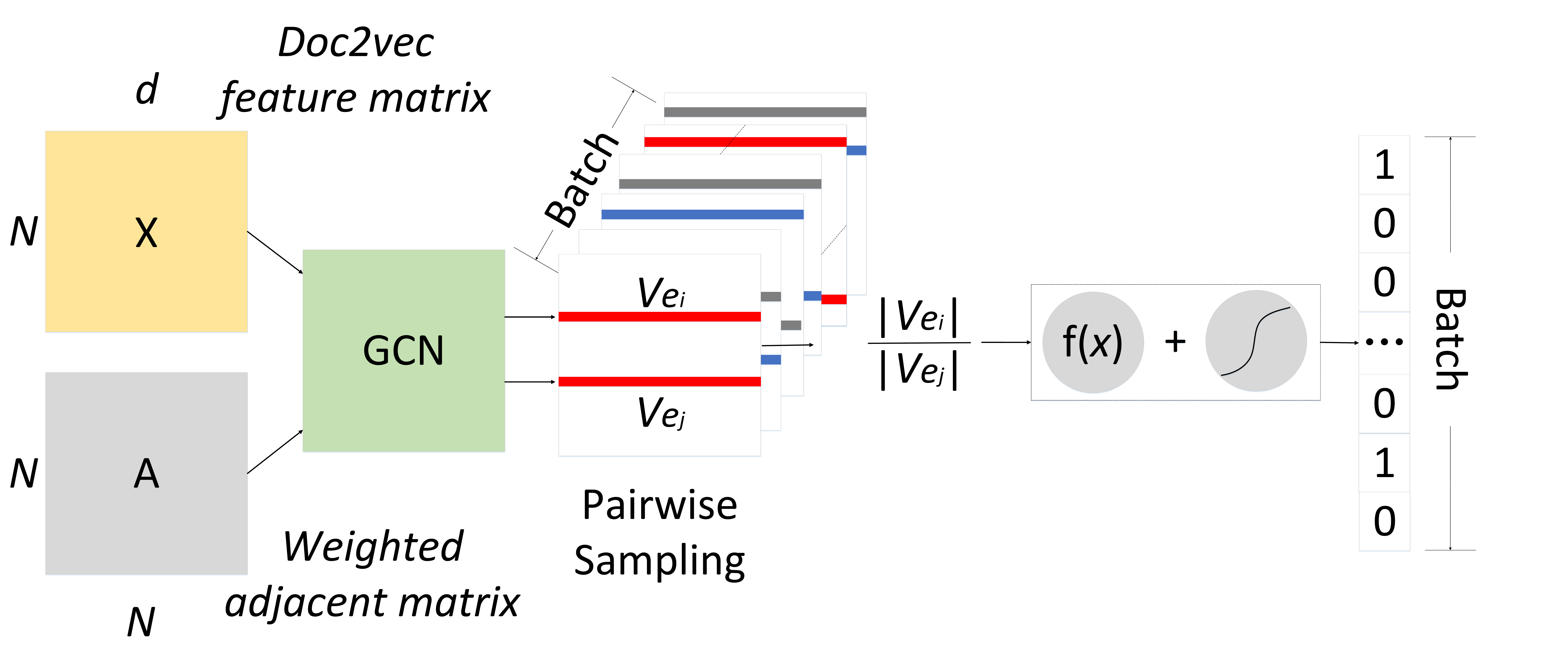}
\caption{An overview of the proposed Pairwise Popularity Graph Convolutional Network (PP-GCN).}\label{fig:pp-gcn}
\end{figure}

However, the real-world social events naturally have two problems of sparsity: the small number of event instances for each classification and a large number of categories.
So, we sample event instances pair and judge whether the pair belongs to one event to train a pairwise GCN model.
As shown in Figure~\ref{fig:pp-gcn}, we present the proposed PP-GCN model.
Before explaining the PP-GCN model, we show how to implement a pairwise sampling to generate training samples.
We assume that if a pair of event instances $e_i$ and $e_j$ belongs to the same event classification, we name the pair $e_i$ and $e_j$ as a \emph{positive-pair} sample.
If a pair of event instances $e_i$ and $e_j$ belongs to two different events classification, we name the pair $e_i$ and $e_j$ as a \emph{negative-pair} sample.
As shown in Figure~\ref{fig:pp-gcn}, if the pair is a {positive-pair} sample, we represent its by two red lines; if the pair is {negative-pair} sample, we use both gray line and blue line to represent it.
After explaining the training samples, we first randomly select $R$ (i.e., 1000) event instances as a preliminary set, then randomly select two event instances for each event instance in the set to form one {positive-pair} sample and one  {negative-pair} sample, and finally we can construct a $2R$ event instance pairs set from training samples.
Here, both the {positive-pair} and {negative-pair} samples are equal to $R$.
Second, we randomly sample the $B$ (i.e., 64) samples from the $2R$ (i.e., 2000) event instance pairs set to form a batch to forward propagation of our proposed model.
Third, the second step is cycled $E$ (i.e., 32) times to form an epoch.
For next epoch, we loop through the above three steps.

However, the above pairwise sampling based GCN model can not guarantee that the model avoids over-fitting during training.
Suppose that any event classification has an average of $r$ event instances, the probability that any event instance selected into a {positive-pair} sample is $\frac{1}{r}$, and the probability of being selected into a {negative-pair} sample is about $\frac{1}{N-r}$.
We note that $\frac{1}{r} \gg \frac{1}{N-r}$ in general.
Obviously, the {negative-pair} samples have more diversity than the {positive-pair} samples.
~\cite{papadopoulos2012popularity} has observed the phenomenon that the connected probability of a sample determines the popularity of it.
Inspired by these observations, we assume that in feature representation learning, the modulus of the learned feature vector is larger if the popularity is greater.
So, the two modulus of learned event instances feature vectors of {positive-pair} will be closer.

For discriminate feature learning of our GCN model, we utilize the popularity of the output event instance feature vector in $Z$ to distinguish different classes.
As shown in the Figure~\ref{fig:pp-gcn}, for any two learned event instance vectors $V_{e_{i}}$ and $V_{e_{j}}$ that satisfy $|V_{e_{i}}|\ge |V_{e_{j}}|$, we employ a ratio of modulus $x=\frac{|V_{e_{i}}|}{|V_{e_{j}}|}$ as the input of a nonlinear mapping function $f(x) = - log(x-1+c)$, where the coefficient $c$ is 0.01 to avoid no upper bound output.
We assume that the ratio of modulus of \emph{positive-pair} will belong to $[1, 2)$.
So, the nonlinear mapping function $f(x)$ can map the above ratio $x$ from [1, 2) to (0, 2], and [2, +$\infty$) to (-$\infty$, 0).
Next, we add a Sigmoid function to map the output of the nonlinear mapping layer to 0 or 1 by a threshold 0.5.
As shown in the Figure~\ref{fig:pp-gcn}, one \emph{positive-pair} or \emph{negative-pair} input sample can only be paired with an output of 0 or 1.
For one batch (64 pairs) samples, our model can generate one batch size ($1\times 64$) of one-zero output vector.
So, we can use a cross entropy function as our model's loss function, and employ the popular stochastic gradient descent (SGD) method to iterate all parameters.
The learned weights $\vec{\omega}$ will be used to measure similarity for any two social event instances.
To verify the avoidance of over-fitting ability of our model, we can perform over 7000 epochs, and observe that the evaluation criteria of the model changes over time in Section~\ref{sec:expe}.

For the testing of any event instance $t$ from the test set of the original $N$ samples, we first assume that there are a total of $C$ event classes in the original $N$ event instances.
Secondly, we calculate the ratio of the modulus of the representation vectors for $t$ and the remaining $N-1$ samples, respectively.
Then, for each event class, we can get a probability that event instance $t$ most likely belongs to it.
If all of the ratios of modulus are 0, the sample itself is a separate event class.
Finally, we select the event class with the highest probability as the test output for the event instance $t$.
Note the fact that the event category of the test set may not be included in the event category of the training set.
After the previous analysis, we can calculate a similarity for any two event instances under the event-HIN and the weights $\vec{\omega}$.
Since our meta-paths have better interpretability, we also can implement a semi-supervised and fine-grained event clustering based on the learned weights of meta-paths and distance-based clustering models.


\section{Experiments}\label{sec:expe}
In this section, we evaluate the proposed PP-GCN model and similarity measure KIES using real surveillance data collected in two enterprise systems.

\begin{table}[h]
\center
\renewcommand{\multirowsetup}{\centering}
	\begin{tabular}{|c|cccc|}
		\toprule
        Datasets& Train& Validation& Test & Class\\
        \midrule
         Tencent& 17,438& 5,813& 5,812& 9,941\\
         Weibo& 6,000& 2,000& 2,000& 5,470\\
		\bottomrule
	\end{tabular}
	\caption{Description of evaluation datasets.\label{tab:Static-data-statistics}}
\end{table}

\subsection{Datasets and Settings}
We select two independent social media platforms, news APP from Tencent (a popular APP for young people) and Sina Weibo (a hybrid of Twitter and Facebook, the Twitter of China and Chinese Social Media), to collect datasets.
Each event instance is a non-repeating social message text.
One event is a set of event instances that contain semantically identical information revolving around a real world incident.
An event always has a specific time of occurrence.
It may involve a group of social users, organizations, participating persons, one or several locations, other types of entities, keywords, topics, etc.
In our work, social events cover a wide variety of types, including a large number of events that occur in the real world and spread on social networks, such as earthquakes, national policies, economic crises, and so on. 
Each event class refers to a unique event.
For example, social media’s tweet about \emph{Tiger Woods winning the 2019 Masters of Golf} is an influential event in the real world and unlike \emph{Patrick Reid’s 2018 Masters of Golf}. 
These are two different events that happen in the real world and belong to different event categories.
The event labels for the Weibo and Tencent datasets are labeled by the outsourcing companies.
Both entities and keywords have been manually extracted for the Tencent dataset.
Note that the anonymized social users and their friend relationships involved in these two datasets are granted by the two companies for scientific research purposes only.
For both of the two datasets, we use 60\% of samples as training set, 20\% of samples as development set and the remaining 20\% of as test set.
The statistics of the two datasets is shown in Table~\ref{tab:Static-data-statistics}.
We can see that the total number of class is large and the number of samples in per class is small.

We conduct the experiments on event detection and event clustering on these two datasets.
The operating system and software platforms are Ubuntu 5.4.0, Tensorflow-gpu (1.4.0) and Python 2.7.
The metrics used to evaluate the performance of event detection are the accuracy and F1 score. 
The metric used to evaluate the performance of event clustering is the normalized mutual information (NMI). 

\subsection{Baseline Methods}\label{sec:baselines}
Since the work of fine-grained social event categorization is relatively small, we briefly describe the baseline methods of text matching and text distance.
For all the baselines, we use the implementations or open source codes of these models released by authors and other researchers, and report the best performance of the results.

\textbf{Support Vector Machine with TF-IDF feature (SVM)}: Support Vector Machine with pair document TF-IDF features is the most classical approach for classification task. 
We extract the TF-IDF features for social messages, and then use the SVM classifier to implement the multi-class event classification.
\textbf{Convolutional Matching Architecture-I (ARC-I)}~\cite{hu2014convolutional}: It encodes text pairs by CNNs, and compares the encoded representations of each text with a MLP.
\textbf{Convolutional Matching Architecture-II (ARC-II)}~\cite{hu2014convolutional}: It builds directly on the interaction space between two texts, and models all the possible combinations of them with 1-D and 2D convolutions.
\textbf{Match by Local and Distributed Representations (DUET)}~\cite{mitra2017learning}: It matches two texts using both local representation and learned distributed representation.
\textbf{Multiple Positional Semantic Matching (MV-LSTM)}~\cite{wan2016deep}: It matches two texts with multiple positional text representations, and aggregates interactions between different positional representations.
\textbf{Convolutional Deep Structured Semantic Models (C-DSSM)}~\cite{shen2014learning}: It learns low-dimensional semantic vectors for input text by CNNs.
\textbf{Deep Structured Semantic Model (DSSM)}~\cite{huang2013learning}: It utilizes a deep neural network to map high-dimensional sparse features into low-dimensional features, and calculates the semantic similarity of the document pair.
\textbf{Siamese Encoded Graph Convolutional Network (SE-GCN)}~\cite{liu2018matching}: It learns vertex representations through a Siamese neural network and aggregates the vertex features though GCNs to generate the document matching.

\textbf{Term Frequency-Inverse Document Frequency (TF-IDF)}: It uses the bag-of-words representation divided by each word’s document frequency.
\textbf{Latent Dirichlet Allocation (LDA)}~\cite{blei2003latent}: is a celebrated generative model for text documents that learns representations for documents as distributions over word topics.
\textbf{Marginalized Stacked Denoising Autoencoder (mSDA)}~\cite{chenmarginalized}: It is a representation learned from stacked denoting autoencoders.
\textbf{Componential Counting Grid (CCG)}~\cite{perina2013documents}: It is a generative model that models documents as a mixture of word distributions and LDA.
\textbf{Word Move Distance (WMD)}~\cite{kusner2015word}: It measures the dissimilarity between two documents as the minimum amount of distance that words of one document need to travel to reach words of another document.
\textbf{Knowledge-driven document similarity measure (KnowSim)}~\cite{wang2016text}: It's also a meta-paths instances based document similarity, and hasn't considered the impacts of social users. The weights of meta-paths are estimated by the Laplacian scores of documents.

\subsection{Performance Analysis}
Table~\ref{tab:event-detection} shows the accuracy and F1-score of different algorithms on the task of event detection in Tencent and Weibo datasets.
Overall, the proposed PP-GCN model consistently and significantly outperforms all baselines in terms of accuracy and F1.
In the Tencent dataset, PP-GCN achieves 13\%–56\% improvements in terms of accuracy and F1 over all baselines.
In the Weibo dataset, PP-GCN achieves 10\%–33\% improvements in terms of accuracy and F1 over all baselines.

\begin{table}[t]
\renewcommand{\multirowsetup}{\centering}
\begin{tabular}{|c|cccc|}
\toprule
\multirow{2}{*}{Algorithms} & \multicolumn{2}{c}{Tencent} & \multicolumn{2}{c}{Weibo}\\
 & Accuracy & F1 & Accuracy & F1\\
\midrule
ARC-I & 0.5384 & 0.4868 &  0.4910 & 0.4857 \\
ARC-II & 0.5437 & 0.3677 &  0.5277 & 0.5137 \\
DUET & 0.5625 & 0.5237 & 0.5397 & 0.5523 \\
DSSM & 0.5808 & 0.6468 & 0.5765 & 0.5411 \\
C-DSSM & 0.6017 & 0.4857 & 0.6170 & 0.5814 \\
MV-LSTM & 0.5562 & 0.6383 & 0.6252 & 0.6613\\
SVM & 0.7581 & 0.7361 &  0.6511 & 0.6268 \\
SE-GCN & 0.7901 & 0.7893 & 0.7063 & 0.7015 \\
\midrule
PP-SE-GCN & 0.8319 & 0.8383 & 0.7317 & 0.7384 \\
PA-GCN & 0.8818 & 0.8801 & 0.7567 & 0.7591 \\
PP-GCN& \bf{0.9252} & \bf{0.9231} & \bf{0.8000} & \bf{0.8134} \\
\bottomrule
\end{tabular}
\caption{\label{tab:event-detection}Accuracy and F1 results of event detection.} \center
\end{table}

The improvements can be attributed to the three characteristics of proposed models.
First, the knowledgeable HIN is better modeling social events than traditional text modeling methods, such as bag-of-words (SVM), N-gram (ARC-I, ARC-II and C-DSSM) and sequence-of-words (MV-LSTM).
Our PP-GCN has improved overall by more than 10\% in the event detection over the SE-GCN model incorporating structural and conceptual semantics.
Second, the combination of KIES based weighted adjacent matrix and Doc2Vec is better for fine-grained event instance representation learning than for feature extraction on text pairs, such as DUET.
Third, the classifier based on the ratio of modulus of generated representations of event instances is better than the traditional pairwise distances.
Here, we replace the regression module of SE-GCN model by our proposed popularity based classifier, named by PP-SE-GCN, and the performances can be improved 3\%-5\% in Tencent and Weibo.
The 10\%-14\% improvements from the SE-GCN to the PP-GCN demonstrate the advantages of knowledgeable HIN modeling and the pairwise popularity based feature learning framework.

\begin{figure}[h]
\center
\includegraphics[width=0.5\textwidth]{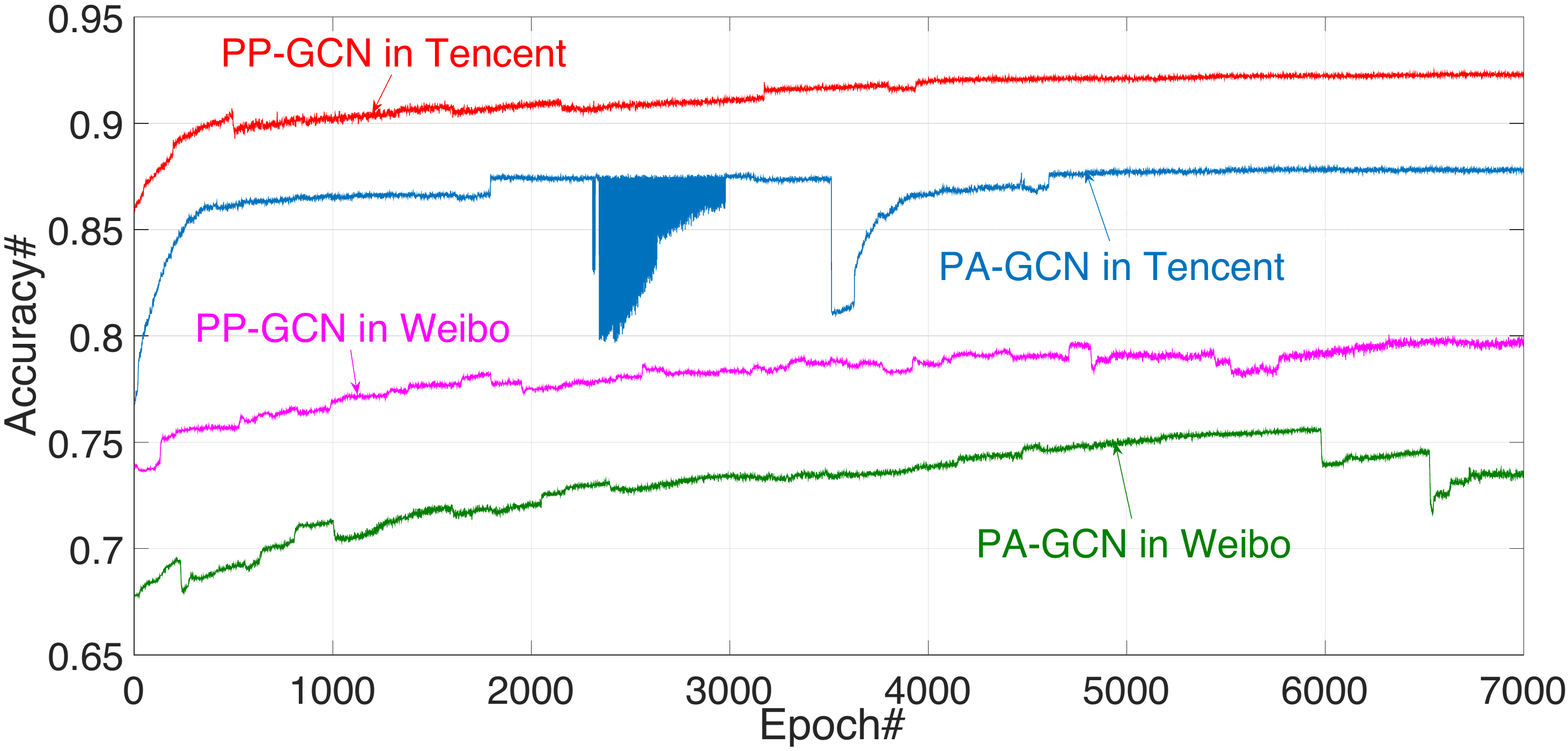}
\caption{Illustration of the Accuracy for PP-GCN and PA-GCN.}\label{fig:gcn_acc}
\end{figure}

\begin{table*}[t]
\center
\renewcommand{\multirowsetup}{\centering}
\begin{tabular}{|c|cccccc|cc|}
\toprule
Datasets& TF-IDF& LDA &  mSDA& CCG & WMD & KnowSim & KIES& KIES(T)\\
\midrule
Tencent& 0.6686 & 0.6979 & 0.7545 & 0.7715 & 0.8166 & 0.8059 & \bf{0.9012} & 0.8937 \\
Weibo & 0.5824 & 0.6014 & 0.6518 & 0.6973 & 0.7261 & 0.7191 & 0.7820  & \bf{0.8041} \\
\bottomrule
\end{tabular}
\caption{\label{tab:event-clustering}NMI results of event clustering.}
\end{table*}

Furthermore, our PP-GCN model can avoid over-fitting in training.
We replace our classifier in PP-GCN by the angle of generated event instances feature vectors based classifier, namely PA-GCN.
In Figure~\ref{fig:gcn_acc}, we visualize the test accuracies of the PP-GAN and PA-GCN in Tencent and Weibo in 7000 epochs.
From Figure~\ref{fig:gcn_acc}, we observe that the overall trend of the accuracies of the PP-GCN model is continuously increasing, but the accuracies of the angle-based PA-GCN model have periodic fluctuations. 
Essentially, the ratio of the modulus between vectors is more stable than the angle in iterations.
Compared to the angle-based classifier, the popularity-based classifier has better ability to learn discriminating and stable event instance feature and prevent overfitting.

One advantage of the proposed PP-GCN compared to other methods is that the weights $\vec{\omega}$ between the meta-paths can be learned according to the event detection task.
Due to the interpretability of the meta-path and similarity measure KIES, the learned weights $\vec{\omega}$ can be utilized in other applications.
Here, we make use of different similarity measures including KIES and other methods discussed in Section~\ref{sec:baselines}, and leverage the popular k-means algorithm to cluster the events.
For the KIES distance metric, we use the two weights $\vec{\omega}$ learned in the event detection tasks of Tencent and Weibo datasets, and then calculate the KIES distances between event instances by Eq.~\ref{eq:event_sim_inst} to implement a semi-supervised fine-grained event clustering.
Note that the test data did not participate in the training when learned the meta-path weights $\vec{\omega}$ in PP-GCN.

As shown in Table~\ref{tab:event-clustering}, our proposed similarity measure KIES achieves the best performances on the two clusters tasks in terms of NMI.
Moreover, among the baselines, the WMD, mSDA and CGG measures have been verified to achieve state-of-the-art effects in text similarity in ~\cite{kusner2015word}.
Compared to other similarity measures, our KIES based k-means method achieves 6\%–24\% improvements in terms of NMI.
We even implement a meta-path weights transfer experiment between Tencent dataset and Weibo dataset.
We see that the performance of the Weibo dataset are improved more than 2\% when employing the weights of the Tencent dataset to the Weibo by the KIES(T) based k-means method, but not the other way around. 
We believe the reason is that Tencent dataset has more applicable meta-path weights by training with manually labeled entities and keywords.
Based on the learned weights and the interpretable distance metric KIES, we have achieved the best performance of semi-supervised event clustering.

\section{Conclusion}
In this paper, we propose a knowledgeable HIN based social event modeling framework, and design a novel pairwise popularity GCN model to learn both meta-paths weights and discriminant event instance representation, and achieves fine-grained social event categorization with state-of-the-art performances.
By using the proposed PP-GCN model, we are able to overcome the problems of large category size and sparse small number of samples per class and preventing overfitting in our tasks.
Experimental results show that our PP-GCN and KIES similarity measure can significantly outperform state-of-the-art baselines methods on two real-world social datasets.
In the future, we plan to study the interpretability of the different importance of meta-paths, and extend our framework to other complex parameter leaning and applications.

\section*{Acknowledgements}
The corresponding author is Jianxin Li. 
This work is supported by NSFC program (No.61872022, No.61421003) and SKLSDE-2018ZX-16.
Yangqiu Song is supported by the Early Career Scheme (ECS, No. 26206717) from Research Grants Council in Hong Kong.
Philip S. Yu is supported by NSF through grants IIS-1526499, IIS-1763325, and CNS-1626432, and NSFC No.61672313.

\bibliographystyle{named}
\bibliography{ms}

\begin{thebibliography}{}

\bibitem[\protect\citeauthoryear{Aggarwal and
  Subbian}{2012}]{aggarwal2012event}
Charu~C Aggarwal and Karthik Subbian.
\newblock Event detection in social streams.
\newblock In {\em SDM}, 2012.

\bibitem[\protect\citeauthoryear{Allan}{2012}]{Allan:2012:TDT:2481012}
James Allan.
\newblock {\em Topic Detection and Tracking: Event-based Information
  Organization}.
\newblock Springer Publishing Company, Incorporated, 2012.

\bibitem[\protect\citeauthoryear{Angel \bgroup \em et al.\egroup
  }{2012}]{angel2012dense}
Albert Angel, Nikos Sarkas, Nick Koudas, and Divesh Srivastava.
\newblock Dense subgraph maintenance under streaming edge weight updates for
  real-time story identification.
\newblock {\em PVLDB}, 2012.

\bibitem[\protect\citeauthoryear{Atefeh and Khreich}{2015}]{atefeh2015survey}
Farzindar Atefeh and Wael Khreich.
\newblock A survey of techniques for event detection in twitter.
\newblock {\em Computational Intelligence}, 31(1):132--164, 2015.

\bibitem[\protect\citeauthoryear{Auer \bgroup \em et al.\egroup
  }{2007}]{Auer:2007:DNW:1785162.1785216}
S\"{o}ren Auer, Christian Bizer, Georgi Kobilarov, Jens Lehmann, Richard
  Cyganiak, and Zachary Ives.
\newblock Dbpedia: A nucleus for a web of open data.
\newblock In {\em ASWC}, 2007.

\bibitem[\protect\citeauthoryear{Becker and
  Gravano}{2011}]{becker2011identification}
Hila Becker and Luis Gravano.
\newblock {\em Identification and characterization of events in social media}.
\newblock Columbia University, 2011.

\bibitem[\protect\citeauthoryear{Blei \bgroup \em et al.\egroup
  }{2003}]{blei2003latent}
David~M Blei, Andrew~Y Ng, and Michael~I Jordan.
\newblock Latent dirichlet allocation.
\newblock {\em Journal of machine Learning research}, 3(Jan):993--1022, 2003.

\bibitem[\protect\citeauthoryear{Blei \bgroup \em et al.\egroup
  }{2010}]{blei2010nested}
David~M Blei, Thomas~L Griffiths, and Michael~I Jordan.
\newblock The nested chinese restaurant process and bayesian nonparametric
  inference of topic hierarchies.
\newblock {\em JACM}, 2010.

\bibitem[\protect\citeauthoryear{Chandola \bgroup \em et al.\egroup
  }{2009}]{chandola2009anomaly}
Varun Chandola, Arindam Banerjee, and Vipin Kumar.
\newblock Anomaly detection: A survey.
\newblock {\em ACM computing surveys (CSUR)}, 2009.

\bibitem[\protect\citeauthoryear{Chen \bgroup \em et al.\egroup
  }{2012}]{chenmarginalized}
Minmin Chen, WUSTL EDU, and Zhixiang~Eddie Xu.
\newblock Marginalized denoising autoencoders for domain adaptation.
\newblock In {\em ICML}, 2012.

\bibitem[\protect\citeauthoryear{Griffiths \bgroup \em et al.\egroup
  }{2004}]{griffiths2004hierarchical}
Thomas~L Griffiths, Michael~I Jordan, Joshua~B Tenenbaum, and David~M Blei.
\newblock Hierarchical topic models and the nested chinese restaurant process.
\newblock In {\em NIPS}, 2004.

\bibitem[\protect\citeauthoryear{Hu \bgroup \em et al.\egroup
  }{2014}]{hu2014convolutional}
Baotian Hu, Zhengdong Lu, Hang Li, and Qingcai Chen.
\newblock Convolutional neural network architectures for matching natural
  language sentences.
\newblock In {\em NIPS}, 2014.

\bibitem[\protect\citeauthoryear{Huang \bgroup \em et al.\egroup
  }{2013}]{huang2013learning}
Po-Sen Huang, Xiaodong He, Jianfeng Gao, Li~Deng, Alex Acero, and Larry Heck.
\newblock Learning deep structured semantic models for web search using
  clickthrough data.
\newblock In {\em CIKM}, 2013.

\bibitem[\protect\citeauthoryear{Ji and Grishman}{2008}]{ji2008refining}
Heng Ji and Ralph Grishman.
\newblock Refining event extraction through cross-document inference.
\newblock {\em Proceedings of ACL}, pages 254--262, 2008.

\bibitem[\protect\citeauthoryear{Kim \bgroup \em et al.\egroup
  }{2009}]{kim2009overview}
Jin-Dong Kim, Tomoko Ohta, Yoshinobu Kano, and Jun'ichi Tsujii.
\newblock Overview of bionlp'09 shared task on event extraction.
\newblock In {\em BioNLP}, 2009.

\bibitem[\protect\citeauthoryear{Kipf and Welling}{2017}]{kipf2017semi}
Thomas~N. Kipf and Max Welling.
\newblock Semi-supervised classification with graph convolutional networks.
\newblock In {\em ICLR}, 2017.

\bibitem[\protect\citeauthoryear{Kusner \bgroup \em et al.\egroup
  }{2015}]{kusner2015word}
Matt Kusner, Yu~Sun, Nicholas Kolkin, and Kilian Weinberger.
\newblock From word embeddings to document distances.
\newblock In {\em ICML}, 2015.

\bibitem[\protect\citeauthoryear{Le and Mikolov}{2014}]{le2014distributed}
Quoc Le and Tomas Mikolov.
\newblock Distributed representations of sentences and documents.
\newblock In {\em ICML}, 2014.

\bibitem[\protect\citeauthoryear{Liu \bgroup \em et al.\egroup
  }{2018}]{liu2018matching}
Bang Liu, Ting Zhang, Di~Niu, Jinghong Lin, Kunfeng Lai, and Yu~Xu.
\newblock Matching long text documents via graph convolutional networks.
\newblock {\em arXiv}, 2018.

\bibitem[\protect\citeauthoryear{Liu \bgroup \em et al.\egroup
  }{2019}]{liu2019event}
Yaopeng Liu, Hao Peng, Jianxin Li, Yangqiu Song, and Xiong Li.
\newblock Event detection and evolution in multi-lingual social streams.
\newblock {\em Frontiers of Computer Science}, 2019.

\bibitem[\protect\citeauthoryear{Mikolov \bgroup \em et al.\egroup
  }{2013}]{mikolov2013distributed}
Tomas Mikolov, Ilya Sutskever, Kai Chen, Greg~S Corrado, and Jeff Dean.
\newblock Distributed representations of words and phrases and their
  compositionality.
\newblock In {\em NIPS}, 2013.

\bibitem[\protect\citeauthoryear{Miller}{1998}]{miller1998wordnet}
George Miller.
\newblock {\em WordNet: An electronic lexical database}.
\newblock MIT press, 1998.

\bibitem[\protect\citeauthoryear{Mitra \bgroup \em et al.\egroup
  }{2017}]{mitra2017learning}
Bhaskar Mitra, Fernando Diaz, and Nick Craswell.
\newblock Learning to match using local and distributed representations of text
  for web search.
\newblock In {\em WWW}, 2017.

\bibitem[\protect\citeauthoryear{Papadopoulos \bgroup \em et al.\egroup
  }{2012}]{papadopoulos2012popularity}
Fragkiskos Papadopoulos, Maksim Kitsak, M~{\'A}ngeles Serrano, Mari{\'a}n
  Bogun{\'a}, and Dmitri Krioukov.
\newblock Popularity versus similarity in growing networks.
\newblock {\em Nature}, 489(7417):537, 2012.

\bibitem[\protect\citeauthoryear{Perina \bgroup \em et al.\egroup
  }{2013}]{perina2013documents}
Alessandro Perina, Nebojsa Jojic, Manuele Bicego, and Andrzej Truski.
\newblock Documents as multiple overlapping windows into grids of counts.
\newblock In {\em NIPS}, 2013.

\bibitem[\protect\citeauthoryear{Ritter \bgroup \em et al.\egroup
  }{2012}]{ritter2012open}
Alan Ritter, Oren Etzioni, Sam Clark, et~al.
\newblock Open domain event extraction from twitter.
\newblock In {\em KDD}, 2012.

\bibitem[\protect\citeauthoryear{Shao \bgroup \em et al.\egroup
  }{2017}]{shao2017efficient}
Minglai Shao, Jianxin Li, Feng Chen, Hongyi Huang, Shuai Zhang, and Xunxun
  Chen.
\newblock An efficient approach to event detection and forecasting in dynamic
  multivariate social media networks.
\newblock In {\em WWW}, 2017.

\bibitem[\protect\citeauthoryear{Shen \bgroup \em et al.\egroup
  }{2014}]{shen2014learning}
Yelong Shen, Xiaodong He, Jianfeng Gao, Li~Deng, and Gr{\'e}goire Mesnil.
\newblock Learning semantic representations using convolutional neural networks
  for web search.
\newblock In {\em WWW}. ACM, 2014.

\bibitem[\protect\citeauthoryear{Shi \bgroup \em et al.\egroup
  }{2017}]{shi2017survey}
Chuan Shi, Yitong Li, Jiawei Zhang, Yizhou Sun, and Philip.S Yu.
\newblock A survey of heterogeneous information network analysis.
\newblock {\em TKDE}, pages 17--37, 2017.

\bibitem[\protect\citeauthoryear{Sun and Han}{2013}]{sun2013mining}
Yizhou Sun and Jiawei Han.
\newblock Mining heterogeneous information networks: a structural analysis
  approach.
\newblock {\em Acm Sigkdd Explorations Newsletter}, 2013.

\bibitem[\protect\citeauthoryear{Wan \bgroup \em et al.\egroup
  }{2016}]{wan2016deep}
Shengxian Wan, Yanyan Lan, Jiafeng Guo, Jun Xu, Liang Pang, and Xueqi Cheng.
\newblock A deep architecture for semantic matching with multiple positional
  sentence representations.
\newblock In {\em AAAI}, 2016.

\bibitem[\protect\citeauthoryear{Wang \bgroup \em et al.\egroup
  }{2016}]{wang2016text}
Chenguang Wang, Yangqiu Song, Haoran Li, Ming Zhang, and Jiawei Han.
\newblock Text classification with heterogeneous information network kernels.
\newblock In {\em Thirtieth AAAI Conference on Artificial Intelligence}, 2016.

\bibitem[\protect\citeauthoryear{Wang \bgroup \em et al.\egroup
  }{2018}]{wang2018unsupervised}
Chenguang Wang, Yangqiu Song, Haoran Li, Ming Zhang, and Jiawei Han.
\newblock Unsupervised meta-path selection for text similarity measure based on
  heterogeneous information networks.
\newblock {\em Data Mining and Knowledge Discovery}, 32(6):1735--1767, 2018.

\bibitem[\protect\citeauthoryear{Xu \bgroup \em et al.\egroup
  }{2017}]{xu2017cn}
Bo~Xu, Yong Xu, Jiaqing Liang, Wanyun Cui, and Yanghua Xiao.
\newblock Cn-dbpedia: A never-ending chinese knowledge extraction system.
\newblock In {\em IEA/AIE}, 2017.

\bibitem[\protect\citeauthoryear{Yu \bgroup \em et al.\egroup
  }{2017}]{yu2017ring}
Weiren Yu, Jianxin Li, Md~Zakirul~Alam Bhuiyan, Richong Zhang, and Jinpeng
  Huai.
\newblock Ring: Real-time emerging anomaly monitoring system over text streams.
\newblock {\em IEEE Transactions on Big Data}, 2017.

\bibitem[\protect\citeauthoryear{Zacks and Tversky}{2001}]{zacks2001event}
Jeffrey~M Zacks and Barbara Tversky.
\newblock Event structure in perception and conception.
\newblock {\em Psychological bulletin}, 2001.

\bibitem[\protect\citeauthoryear{Zhang \bgroup \em et al.\egroup
  }{2014}]{zhang2014part}
Ning Zhang, Jeff Donahue, Ross Girshick, and Trevor Darrell.
\newblock Part-based r-cnns for fine-grained category detection.
\newblock In {\em ECCV}, 2014.

\end{thebibliography}

\end{document}